\documentclass[%
 reprint,
 superscriptaddress,
 nofootinbib,
 amsmath,amssymb,
 aps, prx,
]{revtex4-2}

\usepackage{graphicx}
\usepackage{dcolumn}
\usepackage{bm}

\usepackage{xcolor}
\usepackage{soul}
\usepackage[normalem]{ulem}

\def\NOTE#1{{\textcolor{purple}{#1}}}  
\def\MB#1{{\textcolor{teal}{#1}}}  

\begin{document}

\title{Physics-Constrained Diffusion Model for Synthesis of 3D Turbulent Data}

\author{Tianyi Li}
\author{Michele Buzzicotti}
 \email{Contact author: michele.buzzicotti@roma2.infn.it}
\author{Fabio Bonaccorso}
\author{Luca Biferale}
\affiliation{
 Department of Physics and INFN, University of Rome ``Tor Vergata'', Via della Ricerca Scientifica 1, 00133 Rome, Italy
}

\date{\today}

\begin{abstract}
Synthesizing fully developed three-dimensional turbulent velocity fields remains a long-standing problem in fluid mechanics and an open challenge for generative modeling. 
The difficulty arises from the coexistence of extreme dimensionality, multiscale rough fluctuations and strong intermittency, together with exact physical constraints such as incompressibility and zero-mean momentum.
We propose a physics-constrained diffusion model (PCDM) in which these \emph{a priori} constraints are incorporated directly into the generative dynamics. 
Using rotating turbulence as a stringent benchmark, we show that the proposed framework enables stable and statistically faithful synthesis of inertial-range three-dimensional turbulent velocity fields at medium resolution, accurately reproducing anisotropic energy spectra, intermittency statistics, and physical constraints.
By contrast, standard denoising diffusion probabilistic models without such constraints exhibit multiscale statistical deviations, violations of physical consistency, and substantially slower training convergence.
These findings point to broader implications for generative modeling of high-dimensional complex systems under physical constraints.
\end{abstract}

\maketitle

\section{\label{sec:intro}Introduction}

The synthesis of turbulent flows has a long tradition in fluid mechanics \cite{benzi1993random, juneja1994synthetic, biferale1998mimicking, rosales2006minimal, friedrich2021explicit, shen2024designing, warnecke2025ensemble}, motivated by the need for controlled realizations of complex velocity fields both to advance fundamental understanding and to support modeling applications. Synthetic turbulence has been used as a controlled surrogate to probe statistical aspects of transport, mixing, and intermittency, and as a practical tool in numerical simulations, including the generation of turbulent initial and inflow conditions \cite{jarrin2006synthetic, yousif2023transformer}. 
Despite decades of progress, generating realistic 3D turbulent velocity fields remains a challenging task due to the enormous number of active degrees of freedom distributed across a wide range of scales, strong non-Gaussian intermittency, and strict physical constraints such as incompressibility. A typical instantaneous velocity field is characterized by nontrivial correlations across scales and rough, non-differentiable increments with fractional local H\"older exponent \cite{frisch1996turbulence}.

Recent advances in data-driven generative machine learning, and diffusion models in particular~\cite{ho2020denoising, nichol2021improved, dhariwal2021diffusion}, offer a seemingly promising alternative for turbulence synthesis. Diffusion models have demonstrated remarkable success in learning high-dimensional probability distributions, most notably in image, audio, and video generation, where they are able to produce samples with high perceptual fidelity and statistical realism. 
Motivated by this success, diffusion-based generative models have recently begun to be explored in a variety of scientific applications, including the modeling of nonlinear dynamical systems and probabilistic weather forecasting \cite{gao2024generative, price2025probabilistic}.
From a physical standpoint, these developments suggest that such models may, at least in principle, be capable of capturing the complex multiscale correlations encoded in instantaneous turbulent velocity fields, whose single-time configurations already reflect strong spatial coupling across a wide range of scales and directions.

On the other hand, a typical instantaneous 3D configuration of a turbulent flow requires the capacity to generate velocity fields ranging from moderate resolutions with $O(100^3)$ grid points up to extreme direct numerical simulations at high Reynolds numbers. 
For instance, the current world record for homogeneous and isotropic turbulence (HIT) reaches $32768^3$ grid points \cite{yeung2025gpu, yeung2025small}. 
Generation must be both kinematically and statistically faithful, reproducing physically relevant quantities that span many orders of magnitude across scales due to the power-law scaling of turbulent energy spectra.
The combination of the huge size of the target data and their extreme statistical complexity makes the synthesis of 3D turbulent fields an extreme challenge for generative models.
As a result, the application of diffusion models and other data-driven tools to the generation of synthetic turbulent data has so far remained largely limited to relatively simple configurations, such as 2D smooth Navier-Stokes flows with steep energy spectra \cite{li2024learning}, or to reduced observables extracted from 3D turbulence, including single-particle 1D Lagrangian signals \cite{li2024synthetic, li2024generative, li2026deterministic} and partial 2D cuts of the velocity field \cite{li2023multi}.
Exploratory attempts toward fully 3D turbulent field generation do exist \cite{mohan2020spatio, tretiak2022physics, du2024conditional}, but the statistically faithful instantaneous synthesis of 3D turbulent velocity fields remains largely an open problem.

To address these challenges, we propose a physics-constrained diffusion model in which fundamental physical constraints are incorporated directly into the generative dynamics. 
By focusing on incompressibility and zero-mean momentum, the framework explicitly accounts for these \emph{a priori} constraints of the target velocity-field distribution during both learning and generation. 
This formulation enables the stable and statistically faithful synthesis of inertial-range three-dimensional turbulent velocity fields at medium resolution (up to $256^3$ grid points).
This result highlights the importance of incorporating physical constraints directly into the generative dynamics when learning high-dimensional physical systems. 
In contrast, we show that a direct application of standard diffusion models without such constraints produces velocity fields with multiscale statistical deviations and violations of physical consistency. 
The proposed framework also leads to substantially more stable and efficient training compared with unconstrained diffusion models.

In this work, we use rotating turbulence as a paradigmatic challenge, in which 
(i) strong large-scale organization coexists with intense intermittent small-scale fluctuations in a fully three-dimensional setting (see Fig.~\ref{fig:rotating_turbulence}); and 
(ii) the system is strongly out of equilibrium, with energy transferred simultaneously toward both large and small scales \cite{alexakis2018cascades}. 
In contrast to HIT, where multifractal phenomenology provides a useful statistical framework \cite{frisch1996turbulence}, no comparable phenomenological description currently exists for the anisotropic, multiscale, and strongly non-Gaussian statistics observed in rotating turbulence.
This makes rotating turbulence a stringent benchmark for assessing the ability of generative models to reproduce anisotropic and multiscale turbulent statistics.

{\sc Related work.} 
Related efforts have sought to incorporate physical information into diffusion-based generative frameworks through several strategies.
One representative direction augments the diffusion loss with physics-informed penalty terms or residual constraints, as in physics-informed diffusion models~\cite{bastek2024physics}. 
Such approaches can mitigate local violations of governing equations, but act through auxiliary constraint penalties and leave the underlying diffusion training objective unchanged. 
A complementary line of work enforces constraints during sampling through projection or guidance, including diffusion posterior sampling~\cite{chung2022diffusion} and training-free guidance~\cite{ye2024tfg}. 
These methods have proven effective for inverse and conditional reconstruction problems, but do not address the unconditional synthesis problem considered here, where physically admissible velocity fields must be generated directly from noise.
Recent work has also explored enforcing divergence-free constraints within diffusion models for incompressible flows~\cite{genuist2026divergence}.

The remainder of the paper is organized as follows.
In Sec.~\ref{sec:rot_turb}, we introduce rotating turbulence as a challenging benchmark for generative modeling.
Sec.~\ref{sec:physics_constrained} presents the proposed physics-constrained diffusion model.
Sec.~\ref{sec:results} reports the results of the proposed model.
Finally, Sec.~\ref{sec:conclusion} summarizes the main findings and discusses broader implications.

\section{\label{sec:rot_turb}The Rotating turbulence Challenge}

\begin{figure*}
\includegraphics[width=\textwidth]{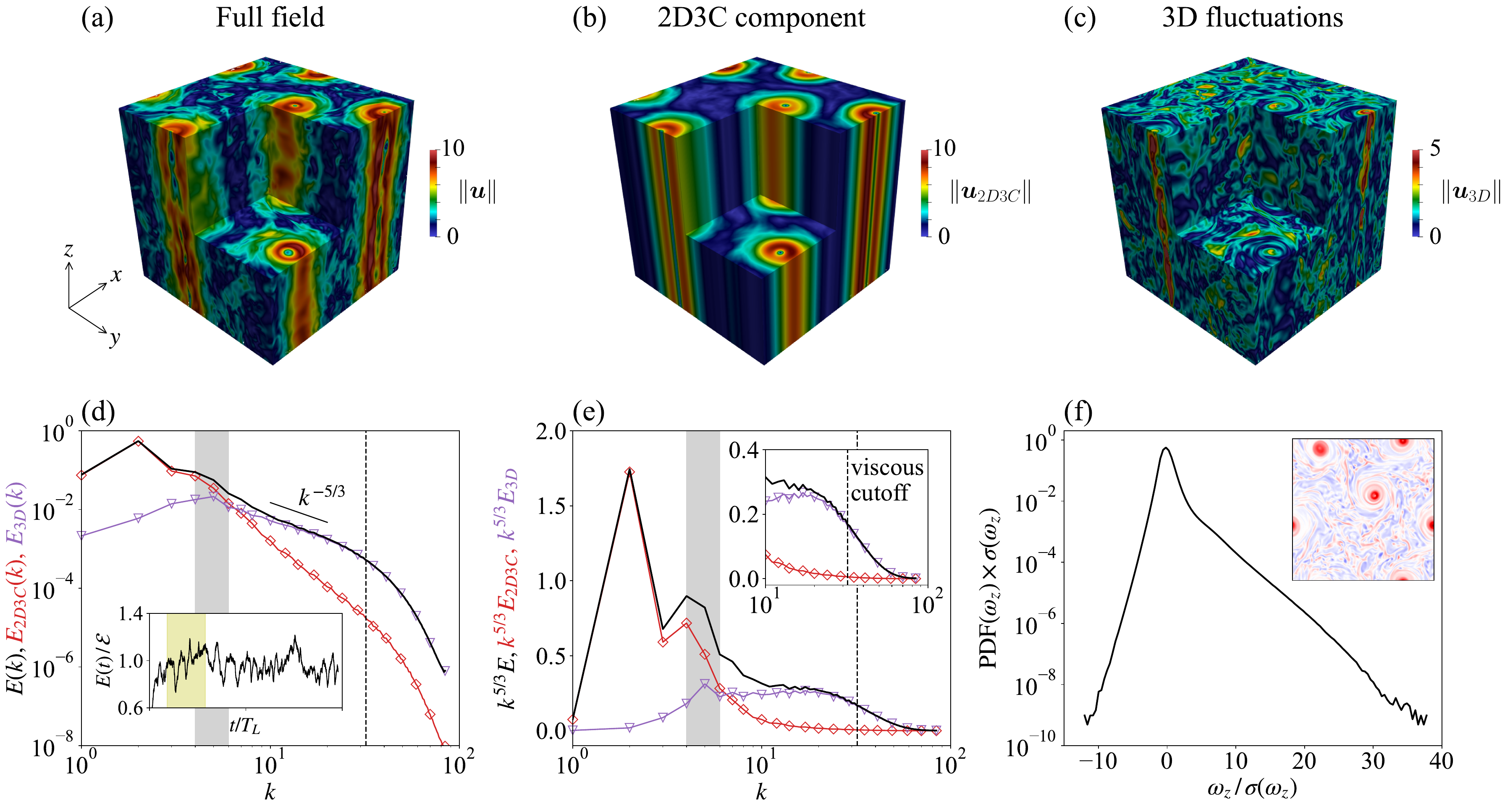}
\caption{\label{fig:rotating_turbulence}
\textbf{Multiscale statistical structure of rotating turbulence.}
(a) Instantaneous three-dimensional rendering of the velocity magnitude $\|\boldsymbol{u}\|$, showing vertically aligned vortical structures induced by strong rotation.
(b) Two-dimensional three-component (2D3C) component of the velocity field obtained from the decomposition of the full field in panel (a), dominated by coherent columnar structures.
(c) Residual three-dimensional (3D) fluctuating field $\boldsymbol{u}_{3D}$ containing small-scale turbulent fluctuations; see Eq.~(\ref{eq:2d3c-3d}) for details.
(d) Total energy spectra for the full field (black solid line), the 2D3C component (red squares), and the 3D fluctuations (purple triangles).
The shaded gray region indicates the forcing band around $k_f$, and the vertical dashed line marks the Kolmogorov wavenumber $k_\nu$.
The inset shows the temporal evolution of the total kinetic energy normalized by its stationary mean; the shaded yellow interval denotes the portion of the simulation used to construct the training dataset (see Appendix~\ref{sec:rot_turb:dataset} for details).
(e) Compensated energy spectra $k^{5/3}E(k)$ highlighting the plateau associated with inertial-range scaling and the viscous cutoff (inset).
(f) Probability density function (PDF) of the vertical vorticity component $\omega_z$, normalized by its standard deviation $\sigma(\omega_z)$, revealing pronounced non-Gaussian tails associated with intermittency.
The inset shows a two-dimensional contour of $\omega_z$ in a plane perpendicular to the rotation axis, with red and blue indicating intense positive and negative vortical structures, respectively.
}
\end{figure*}

The dataset used for training is extracted from a high-resolution direct numerical simulation (DNS) of rotating turbulence, a paradigmatic 3D turbulent configuration of key importance in many industrial and geophysical contexts \cite{godeferd2015structure, buzzicotti2018energy, pedlosky2013geophysical, alexakis2018cascades} governed by the incompressible Navier-Stokes equations in a frame with uniform rotation, $\boldsymbol{\Omega}$: 
\begin{equation}\label{eq:NS_rot1}
    \frac{\partial \boldsymbol{u}}{\partial t}
    + \boldsymbol{u}\cdot\nabla \boldsymbol{u}
    + 2\boldsymbol{\Omega}\times\boldsymbol{u}
    = -\nabla \tilde{p}
    + \nu \Delta \boldsymbol{u}
    + \boldsymbol{f}.
\end{equation}
Details of the numerical setup and data preparation are provided in Appendix~\ref{app:simulation}. 
The resulting velocity field $\boldsymbol{u}(x,y,z,t)$ represents a fully developed turbulent state with strong rotational effects (see Fig.~\ref{fig:rotating_turbulence}(a) for a 3D rendering), evolving chaotically in time (inset of panel (d)), and posing a fundamental challenge for generative modeling. 
In particular, it simultaneously involves an extremely large number of active degrees of freedom, organised in quasi-2D coherent vortical structures (panel (b)) and a strongly anisotropic 3D turbulent background (panel (c)), with pronounced multiscale fluctuations as shown by the power-law energy spectra (panels (d,e)) and strong non-Gaussianity evidenced by the presence of fat tails in the distribution of vorticity (panel (f)).

The first source of complexity arises from the high dimensionality of the flow, involving a large number of dynamically active degrees of freedom distributed across scales. This is quantified by the kinetic energy spectrum (see Fig.~\ref{fig:rotating_turbulence}(d) and Appendix~\ref{app:spectral_definitions}), which exhibits a power-law behaviour, $E(k) \sim k^{-5/3}$, in the wavenumber range $k_f < k < k_\nu$, bounded by the forcing scale $k_f$ and the viscous cutoff $k_\nu$.
A second source of complexity stems from the strong anisotropy induced by rotation, manifested by the coexistence of two-dimensional three-component (2D3C) vortical structures in the direction parallel to rotation and a strongly fluctuating 3D background (see Fig.~\ref{fig:rotating_turbulence}(b,c)). Details on the 2D3C/3D decomposition are provided in Sec.~\ref{sec:results}.
The corresponding energy spectra of the 2D3C component, $E_{2D3C}(k)$, and of the 3D fluctuating contribution, $E_{3D}(k)$, shown in Fig.~\ref{fig:rotating_turbulence}(d,e), show that the 2D3C velocity components dominate at large scales, while the 3D velocity fluctuations are strongly suppressed and become significant only at smaller scales, signaling the gradual recovery of fully three-dimensional dynamics.
The spectrum extends up to the Kolmogorov wavenumber $k_\nu \approx 32$ and decays exponentially beyond that due to viscous dissipation. As a result, the dynamically relevant scales of the flow are well captured by coarse-graining the data on a cube of size $64^3$.

Third, the flow also exhibits strong intermittency.
This is reflected in the probability density function (PDF) of the vertical vorticity component,
$\omega_z = (\nabla \times \boldsymbol{u})_z$,
shown in Fig.~\ref{fig:rotating_turbulence}(f),
which exhibits pronounced non-Gaussian tails associated with intense,
spatially localized vortical structures.
Beyond single-point statistics, intermittency manifests across scales
and is strongly shaped by the anisotropic organization induced by rotation.
These effects will be quantified in Sec.~\ref{sec:intermittency} using scale-dependent flatness
based on transverse velocity increments,
with separate contributions from 2D3C and 3D fluctuations.

Beyond these statistical challenges, the velocity field is subject to strict physical constraints, including incompressibility and zero-mean momentum, which must be satisfied exactly and pose additional challenges for any generative modeling approach.

In the present context, we denote each instantaneous three-dimensional turbulent velocity field of the ground truth DNS by
$$
\mathcal{V}_0 \equiv \{\boldsymbol{u}(\boldsymbol{x}) \,|\, \boldsymbol{x} \in \mathcal{D}\},
$$
where $\mathcal{D}$ is the spatial domain. The velocity field satisfies incompressibility and zero-mean momentum:
\begin{equation}
\nabla \cdot \boldsymbol u = 0 \quad \forall\,\boldsymbol x \in \mathcal D,
\qquad
\langle \boldsymbol u \rangle = \boldsymbol 0,
\label{eq:constraints}
\end{equation}
where $\langle\cdot\rangle$ denotes the spatial average over $\mathcal D$.
Both constraints can be enforced by projecting the velocity field in Fourier space as
\begin{equation}
\widehat{\boldsymbol u}(\boldsymbol k)
\to \left( \mathbf I - \frac{\boldsymbol k\boldsymbol k^\top}{|\boldsymbol k|^2} \right)
\widehat{\boldsymbol u}(\boldsymbol k),
\qquad 
\widehat{\boldsymbol u}(\boldsymbol 0) \to \boldsymbol 0,
\label{eq:projector_fourier}
\end{equation}
where $\widehat{(\cdot)}$ denotes the Fourier transform and $\boldsymbol k$ is the wavevector. 
We denote the projection of a generic 3D velocity field configuration $\mathcal V_0$ using Eq.~(\ref{eq:projector_fourier}) by $\mathcal P(\mathcal V_0)$.
Different implementations of these physical constraints are compared in Sec.~\ref{sec:robustness}.

\section{\label{sec:physics_constrained}Physics-constrained diffusion model}

In this section, we propose a physics-constrained diffusion model (PCDM), obtained by a suitable modification of the standard denoising diffusion probabilistic model (DDPM) \cite{sohl2015deep,ho2020denoising}, whose objective is to approximate the distribution of turbulent flow states $q(\mathcal V_0)$.
This formulation naturally extends to any physical constraint representable as a linear projection in state space; in the present work, we focus on the constraints in Eq.~(\ref{eq:constraints}).

\begin{figure*}
\includegraphics[width=0.9\textwidth]{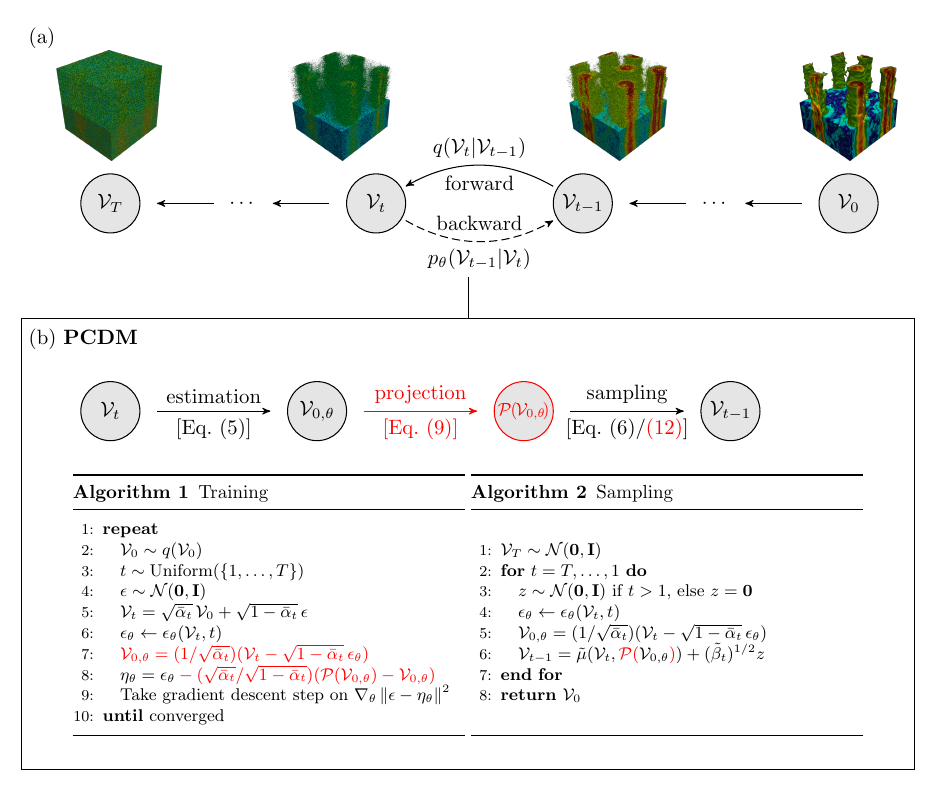}
\caption{\label{fig:ddpm_schematic}
\textbf{Diffusion framework with physics-constrained backward dynamics (PCDM).}
(a) Forward and backward diffusion processes for a 3D turbulent velocity field.
The forward process $q(\mathcal{V}_t|\mathcal{V}_{t-1})$ progressively injects Gaussian noise until the distribution approaches an isotropic Gaussian at $\mathcal{V}_T$, while the learned backward process $p_\theta(\mathcal{V}_{t-1}|\mathcal{V}_t)$ reverses this procedure to generate physically consistent velocity fields.
(b) Physics-constrained backward transition in the PCDM.
The neural network predicts the noise $\epsilon_\theta$, from which a clean estimate for the final generative step $t=0$, $\mathcal{V}_{0,\theta}$, is obtained.
This estimated 3D field is projected onto the physically admissible subspace, enforcing incompressibility and zero-mean momentum, yielding $\tilde{\mathcal{V}}_{0,\theta}$.
A sample $\mathcal{V}_{t-1}$ is then drawn from the Gaussian distribution defined by the reverse transition in Eq.~(\ref{eq:pcdm_reverse_process}), and the procedure is repeated until the final sample $\mathcal{V}_0$ is obtained.
The pseudocode below summarizes the training and sampling procedures of the PCDM.
Removing the projection-based correction (highlighted in red in both the schematic and the algorithms) reduces the method to the standard DDPM formulation.
}
\end{figure*}

The DDPM framework consists of a pair of Markov processes, as illustrated schematically in Fig.~\ref{fig:ddpm_schematic}(a): a fixed forward diffusion process, $q(\mathcal{V}_t | \mathcal{V}_{t-1})$, that progressively corrupts the data with noise, and a learned reverse process, $p_\theta(\mathcal{V}_{t-1} | \mathcal{V}_{t})$, that aims to recover the data starting from independent Gaussian samples.
The forward process generates a sequence of increasingly noisy latent variables
$\mathcal{V}_1, \ldots, \mathcal{V}_T$ by injecting Gaussian noise into a clean sample
$\mathcal{V}_0 \sim q(\mathcal{V}_0)$ according to
\begin{equation}
\mathcal{V}_t
= \sqrt{\bar{\alpha}_t}\,\mathcal{V}_0
+ \sqrt{1-\bar{\alpha}_t}\,\epsilon,
\qquad
\epsilon \sim \mathcal{N}(\boldsymbol{0}, \mathbf{I}),
\label{eq:reparameterization}
\end{equation}
where $\bar{\alpha}_t = \prod_{i=1}^t (1- \beta_i)$ is defined in terms of a prescribed noise schedule $\{\beta_t\}_{t=1}^T$ (see Appendix~\ref{app:ddpm_details} for details).
For a sufficiently large $T$ and an appropriate noise schedule, the distribution of the terminal state
$\mathcal{V}_T$ approaches an isotropic Gaussian,
$q(\mathcal{V}_T) \approx \mathcal{N}(\boldsymbol{0}, \mathbf{I})$.

The reverse (or backward) generative process aims to sample new data 
$\mathcal V_0$ starting from Gaussian noise through the learned reverse transitions 
$p_\theta(\mathcal V_{t-1}|\mathcal V_t)$ parameterized by a neural network.
DDPMs train the neural network to predict the Gaussian noise added during the forward diffusion process, 
$\epsilon_\theta(\mathcal V_t,t)\approx\epsilon$, yielding at each time step an estimate of the clean field
\begin{equation}
\mathcal V_{0,\theta}(\mathcal V_t,t)
=\frac{1}{\sqrt{\bar\alpha_t}}
\left(\mathcal V_t-\sqrt{1-\bar\alpha_t}\,\epsilon_\theta(\mathcal V_t,t)\right).
\label{eq:x0_prediction}
\end{equation}
Following \cite{ho2020denoising}, the reverse transition is parameterized as a learnable Gaussian distribution:
\begin{equation}
p_\theta(\mathcal V_{t-1}|\mathcal V_t)
= 
\mathcal{N}\big(
\mathcal{V}_{t-1};
\tilde{\mu}_t(\mathcal{V}_t, \mathcal V_{0,\theta}),
\tilde{\beta}_t \mathbf{I}
\big),
\label{eq:reverse_process_definition}
\end{equation}
where the mean $\tilde{\mu}_t$ and the variance $\tilde{\beta}_t$ are explicitly defined in Appendix~\ref{app:ddpm_details}.

\subsection{Enforcing Physical Constraints}

While standard DDPMs learn the distribution $q(\mathcal V_0)$ in a purely data-driven manner, 
the velocity fields considered here are known \emph{a priori} to satisfy the constraints in Eq.~(\ref{eq:constraints}).
Within the reverse diffusion dynamics, the network predicts a clean-field estimator 
$\mathcal V_{0,\theta}$ defined in Eq.~(\ref{eq:x0_prediction}), corresponding to the posterior mean of the clean velocity field conditioned on the noisy state $\mathcal V_t$:
\begin{equation}
\mathcal V_{0,\theta} \approx \mathbb E[\mathcal V_0 | \mathcal V_t]
= \int \mathcal{V}_0 \, p(\mathcal V_0 | \mathcal V_t) \, d\mathcal{V}_0,
\label{eq:posterior_mean}
\end{equation}
as implied by Tweedie's formula for Gaussian denoising \cite{chung2022diffusion}.
The estimate $\mathcal V_{0,\theta}$ is not necessarily compliant with the constraints satisfied by the ground-truth velocity fields, whereas the latter satisfy the relations
\begin{equation}
\mathbb E[\mathcal V_0 | \mathcal V_t]
= \mathbb E[\mathcal P(\mathcal V_0) | \mathcal V_t]
= \mathcal P\!\left(\mathbb E[\mathcal V_0 | \mathcal V_t]\right).
\label{eq:proj_cond}
\end{equation}

To enforce condition~(\ref{eq:proj_cond}) for the estimator in Eq.~(\ref{eq:posterior_mean}), we project it onto the physically constrained manifold:
\begin{equation}
\mathcal V_{0,\theta} \rightarrow \mathcal P(\mathcal V_{0,\theta}).
\end{equation}
Replacing $\mathcal V_{0,\theta}$ with its projected counterpart
$\mathcal P(\mathcal V_{0,\theta})$ in Eq.~(\ref{eq:x0_prediction})
yields a corrected noise term $\eta_\theta$:
\begin{equation}
\label{eq:x0_projected}
\mathcal{P}(\mathcal V_{0,\theta})
=\frac{1}{\sqrt{\bar\alpha_t}}
\left(\mathcal V_t-\sqrt{1-\bar\alpha_t}\,\eta_\theta(\mathcal V_t,t)\right),
\end{equation}
given by
\begin{equation}
\eta_\theta
= \epsilon_\theta 
- \frac{\sqrt{\bar{\alpha}_t}}{\sqrt{1-\bar{\alpha}_t}}
\left( \mathcal P(\mathcal V_{0,\theta}) - \mathcal V_{0,\theta} \right).
\label{eq:eps_projection}
\end{equation}
Note that in Eq.~(\ref{eq:x0_projected}) neither the noisy realization $\mathcal V_t$
nor the noise $\eta_\theta$ are required to be divergence-free, whereas the posterior
estimate of the clean velocity field is. In this way we enforce consistency
with the physical constraints through a correction of the predicted noise, without imposing
unnecessary constraints on the noisy intermediate states.
Finally, the reverse transition in Eq.~(\ref{eq:reverse_process_definition}) becomes
\begin{equation}
p_\theta(\mathcal V_{t-1}|\mathcal V_t)
= 
\mathcal{N}\big(
\mathcal{V}_{t-1};
\tilde{\mu}_t(\mathcal{V}_t, \mathcal P(\mathcal V_{0,\theta})),
\tilde{\beta}_t \mathbf{I}
\big).
\label{eq:pcdm_reverse_process}
\end{equation}

As usual, the PCDM model is trained by minimizing the mean-squared error between the true noise injected in the forward process
$\epsilon$ and the corrected prediction $\eta_\theta$,
\begin{equation}
\mathcal{L}_\mathrm{PCDM}
= \mathbb{E}_{t,\mathcal{V}_0,\epsilon}
\left[
\|\epsilon-\eta_\theta(\mathcal{V}_t,t)\|^2
\right],
\label{eq:loss_pcdm}
\end{equation}
where $t$ is sampled uniformly from $\{1,\ldots,T\}$,
$\mathcal V_0 \sim q(\mathcal V_0)$, and $\epsilon \sim \mathcal N(\mathbf 0,\mathbf I)$.
This preserves the standard DDPM denoising objective and therefore remains equivalent 
to maximizing the variational lower bound on the data log-likelihood~\cite{sohl2015deep,ho2020denoising}, 
while explicitly enforcing the \emph{a priori} physical constraints.
A schematic illustration of the PCDM reverse transition, together with the pseudocode for training and sampling, is provided in Fig.~\ref{fig:ddpm_schematic}(b).

Before discussing the superiority of our PCDM over standard DDPM approaches, we note that during the preparation of this manuscript we became aware of a recent preprint~\cite{genuist2026divergence}, which investigates several diffusion-based approaches to enforcing divergence-free constraints. 
In particular, their ``div\_free\_denoiser'' model (Table~1 therein) employs a projection-based correction within the denoising step that is similar to our approach. 
We emphasize that our application, presented in the next section, addresses an open problem in generative modeling of 3D rough and anisotropic turbulent flows. This represents a substantial departure from the proof-of-concept setting considered in~\cite{genuist2026divergence}, where the target fields were obtained from the temporal evolution of decaying, smooth, and isotropic 2D turbulent flows.

\section{\label{sec:results}Generation of 3D turbulence with PCDM}

In this section we present the main results of the PCDM framework for the generation of three-dimensional fully developed turbulent flows, an open challenge for data-driven approaches. 
We first examine the spectral accuracy and training convergence of the model. 
We then analyze multiscale intermittency to assess whether the model faithfully reproduces the non-Gaussian, scale-dependent features of turbulent fluctuations. 
Next, we validate the physical constraints satisfied by the generated fields. 
Finally, we investigate the robustness of the framework across different constraint implementations, showing convergence toward a consistent physical representation.

As shown in many previous studies of rotating turbulence, it is useful to decompose the velocity field into slow and fast manifolds, connected to the presence of inverse and direct energy cascades (see \cite{alexakis2018cascades} and references therein). 
The slow manifold is associated with quasi-coherent two-dimensional vortical structures (the 2D3C component), while the fast manifold corresponds to the remaining 3D fluctuations (see also Fig.~\ref{fig:rotating_turbulence}(a-c) for a graphical visualization):
\begin{equation}
\boldsymbol{u}(x,y,z) = \boldsymbol{u}_{2D3C}(x,y) + \boldsymbol{u}_{3D}(x,y,z),
\label{eq:2d3c-3d}
\end{equation}
where the 2D3C field is defined as the average of the velocity field along the rotation axis,
\begin{equation}
\boldsymbol{u}_{2D3C}(x,y) = \frac{1}{L_0}\int_0^{L_0} \boldsymbol{u}(x,y,z)\,dz,
\end{equation}
where $L_0$ denotes the domain size. 
The field $\boldsymbol{u}_{3D}$ denotes the three-dimensional fluctuating component.

\subsection{\label{sec:spectra}Spectral accuracy and training convergence}

To assess the performance of the proposed framework, we compare the PCDM with two baseline approaches:
(i) a standard DDPM implementation, corresponding to the scheme of Fig.~\ref{fig:ddpm_schematic} without the physics-constrained operations highlighted in red, referred to as DDPM-std; and
(ii) a progressively trained variant of the DDPM (DDPM-prog), designed to alleviate the optimization difficulty associated with the extremely high dimensionality of the full 3D velocity field (see Appendix~\ref{app:progressive}).

Fig.~\ref{fig:training_challenges}(a) shows that the PCDM reproduces the target energy spectra with high fidelity, matching the DNS reference across all scales for both the 2D3C and the 3D components. 
The spectral ratios shown in panels (b,c) demonstrate a clear advantage of the PCDM over the two baseline approaches. 
DDPM-std exhibits pronounced scale-dependent deviations, while DDPM-prog improves the agreement with DNS but retains noticeable deviations over a broad range of scales. 
In contrast, the PCDM provides the closest agreement with the DNS spectra across nearly the entire resolved range of scales.
\begin{figure*}
\includegraphics[width=0.8\textwidth]{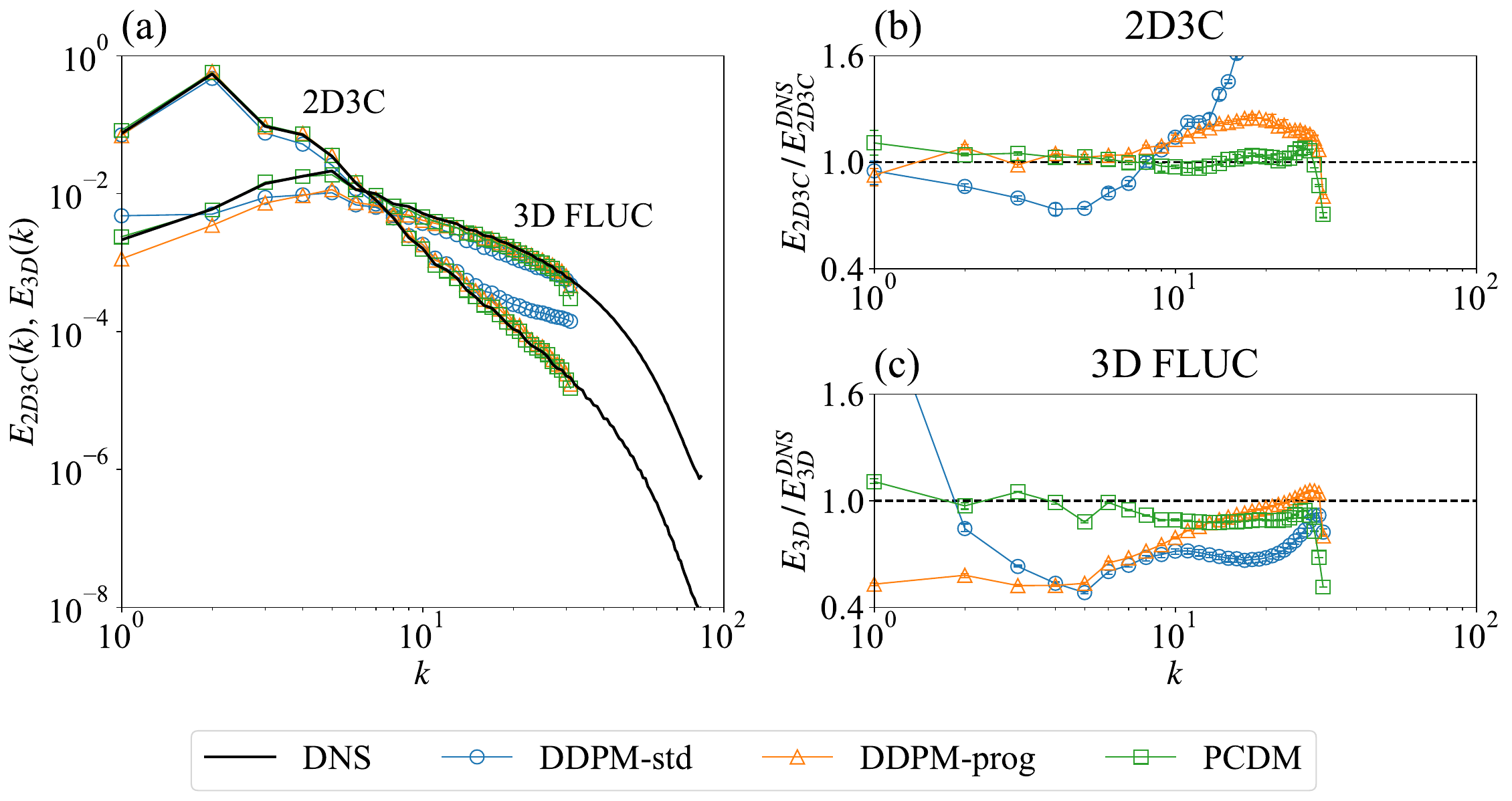}
\caption{\label{fig:training_challenges}
\textbf{Multiscale spectral diagnostics for DDPM-std, DDPM-prog, and the PCDM.}
Here DDPM-std and DDPM-prog denote the standard and progressively trained variants, respectively.
(a) Energy spectra of the 2D3C component and the three-dimensional fluctuating field (3D FLUC), with the DNS reference (black solid line).
(b) Ratio of generated to DNS spectra for the 2D3C component.
(c) Same as (b), but for the 3D fluctuating component.
Error bars are of the order of the symbol size and indicate variability estimated over three independent batches, each consisting of 200 generated fields.
}
\end{figure*}

\begin{figure*}
\includegraphics[width=0.82\textwidth]{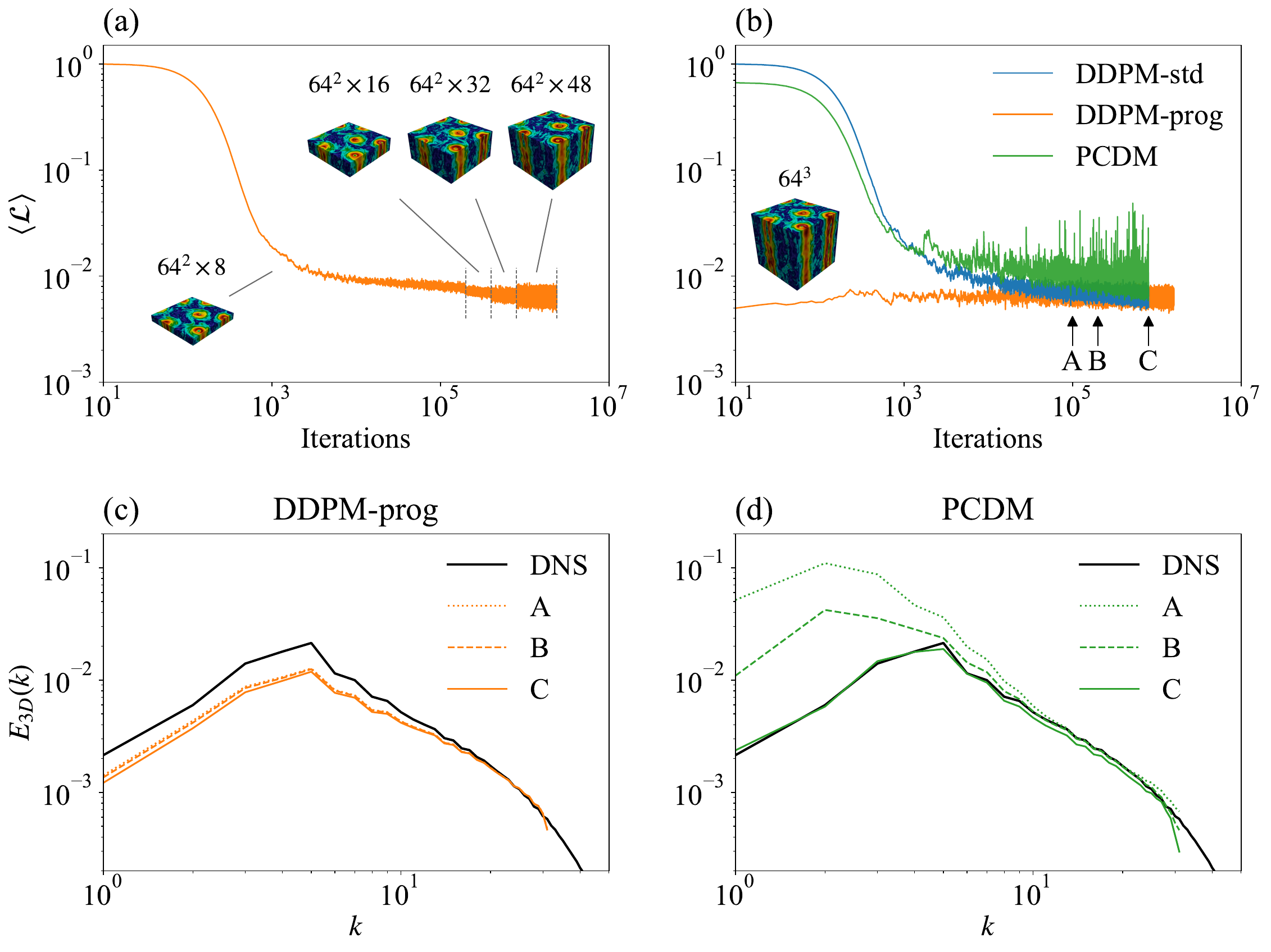}
\caption{\label{fig:training_convergence}
\textbf{Training dynamics and convergence.}
(a) Pretraining loss of DDPM-prog, where the input domain is expanded from a thin slab toward progressively thicker domains.
(b) Training loss of DDPM-std, DDPM-prog, and the PCDM on the full $64^3$ dataset.
Three representative iterations, labeled A, B, and C, are selected from panel (b) to assess convergence.
(c) Energy spectra of the 3D fluctuating component (3D FLUC) generated by DDPM-prog at iterations A, B, and C, compared with the DNS reference (black).
(d) Same as (c), but for the PCDM.
While both models exhibit progressive convergence during training, only the PCDM converges to the DNS spectrum across scales; DDPM-std exhibits persistent spectral fluctuations and is therefore not shown.
}
\end{figure*}

Notably, Fig.~\ref{fig:training_convergence} highlights the superior training behavior of the PCDM compared with the two baseline approaches.
In addition to its improved spectral accuracy, the PCDM converges significantly faster than the unconstrained DDPMs.
In practice, the PCDM reduces the required training time by roughly a factor of five compared with DDPM-prog (see Appendix~\ref{app:training} for details on the computational cost).
Panel (a) shows the pretraining stage of DDPM-prog, where the input domain is gradually expanded from a thin slab toward progressively thicker three-dimensional domains (see Appendix~\ref{app:progressive}).
Panel (b) reports the training loss of DDPM-std, DDPM-prog, and the PCDM during training on the full $64^3$ dataset.
Here $\langle \mathcal{L} \rangle$ denotes the batch-averaged diffusion training loss defined in Eq.~(\ref{eq:loss_pcdm}).
To further assess convergence of the generated fields, three representative training iterations labeled A, B, and C are selected from panel (b).
Panels (c) and (d) report the corresponding energy spectra of the 3D fluctuating component generated by DDPM-prog and the PCDM, respectively.
While DDPM-prog stabilizes after a long pretraining phase, it does not fully recover the DNS spectral behavior across scales.
In contrast, the PCDM progressively converges toward the DNS spectrum during training.

\subsection{\label{sec:intermittency}Intermittency and multiscale statistical fidelity}

Following previous studies \cite{biferale2016coherent}, we characterize multiscale deviations from Gaussian statistics through the analysis of transverse velocity increments measured in the plane normal to the rotation axis,
\begin{equation}
\delta u_\perp(\boldsymbol{r}) =
\left[ \boldsymbol{u}(\boldsymbol{x}+\boldsymbol{r}) - \boldsymbol{u}(\boldsymbol{x}) \right]
\cdot \hat{\boldsymbol{t}},
\end{equation}
where the separation vector $\boldsymbol{r}$ lies in this plane and $\hat{\boldsymbol{t}}$ is orthogonal to both $\boldsymbol{r}$ and the rotation axis.
Assuming isotropy within the plane normal to the rotation axis, the fourth-order flatness is defined as
\begin{equation}
K_\perp^{(4)}(r) =
\frac{\langle [\delta u_\perp(r)]^4 \rangle}
     {\langle [\delta u_\perp(r)]^2 \rangle^2},
\end{equation}
with larger values indicating stronger intermittency.

\begin{figure*}
\includegraphics[width=0.8\textwidth]{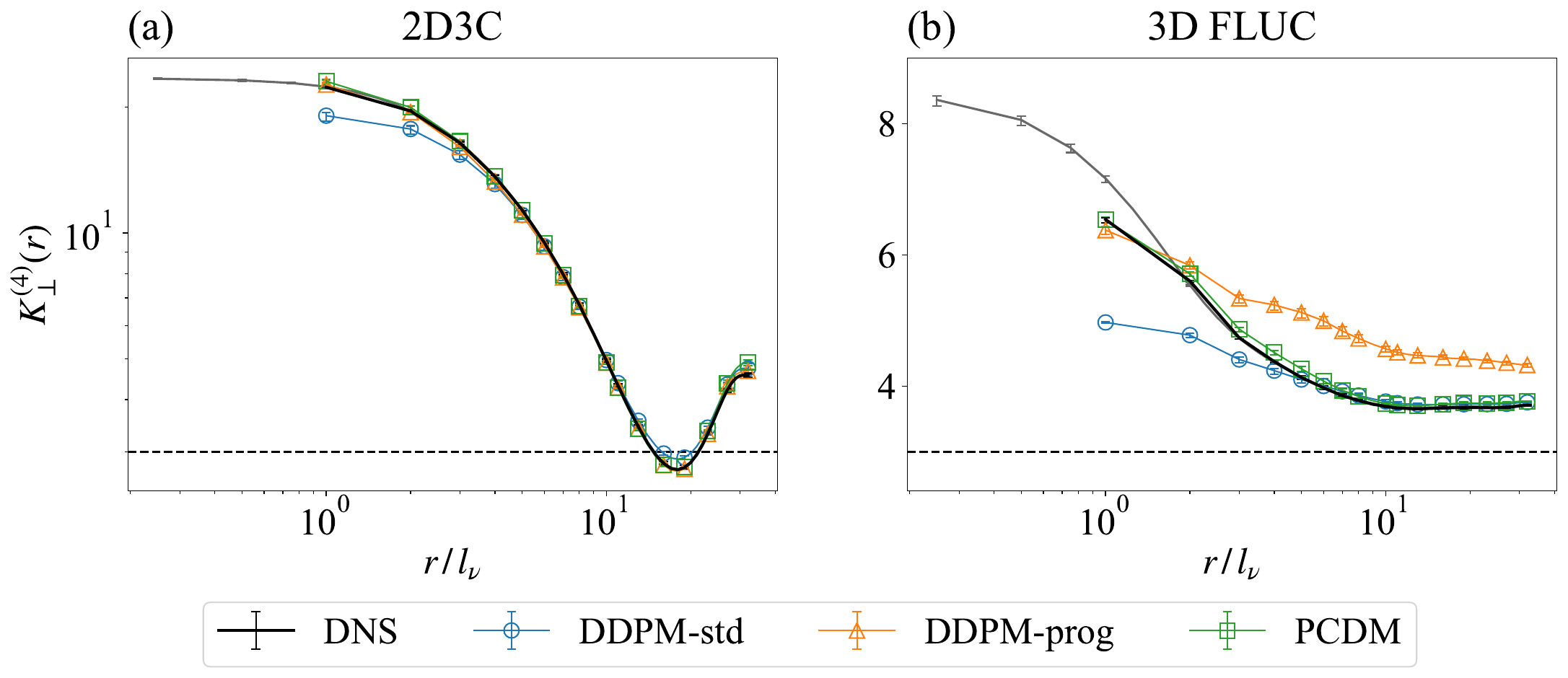}
\caption{\label{fig:flatness}
\textbf{Scale-dependent fourth-order flatness.}
Fourth-order flatness $K_\perp^{(4)}(r)$ of transverse velocity increments,
shown as a function of the separation $r/l_\nu$, where $l_\nu$ denotes the Kolmogorov scale,
measured in the plane normal to the rotation axis,
for the reference DNS ($64^3$), DDPM-std, DDPM-prog, and the PCDM.
Results are reported separately for
(a) the slow 2D3C component and
(b) the fast, fully three-dimensional fluctuating component (3D FLUC).
The dashed horizontal line indicates the Gaussian reference value. 
The gray curve corresponds to the full-resolution DNS data ($256^3$; see Appendix~\ref{sec:rot_turb:dataset}).
}
\end{figure*}

Figure~\ref{fig:flatness} shows the scale-dependent fourth-order flatness of the 2D3C and 3D fluctuating components for the reference DNS and the generated fields. All statistics are evaluated at separations $r/l_\nu$ within the inertial-range scales resolved in the $64^3$ data, where $l_\nu = L_0/k_\nu$ is the Kolmogorov length scale.
As shown, DDPM-std exhibits systematic, scale-dependent discrepancies in both components: for the 2D3C field, the flatness is underestimated at small scales and overestimated in the intermediate-scale dip, while for the 3D fluctuating component, the flatness is captured at large scales but systematically underestimated at small separations. Progressive training (DDPM-prog) improves the agreement for the 2D3C component, yielding a flatness profile close to the DNS reference, consistent with the staged exposure of the model to increasing complexity along the rotation axis. 
In contrast, for the 3D fluctuating component, 
DDPM-prog exhibits large discrepancies across all scales, indicating that the 3D statistics are not captured by progressive training starting from thin quasi-2D slabs.
Conversely, our PCDM accurately reproduces the DNS flatness for both the 2D3C and 3D components across all scales. 
This result shows that the PCDM-generated fields correctly reproduce the intermittent features of both the slow and fast components in rotating turbulence.

\begin{figure*}
\includegraphics[width=0.82\textwidth]{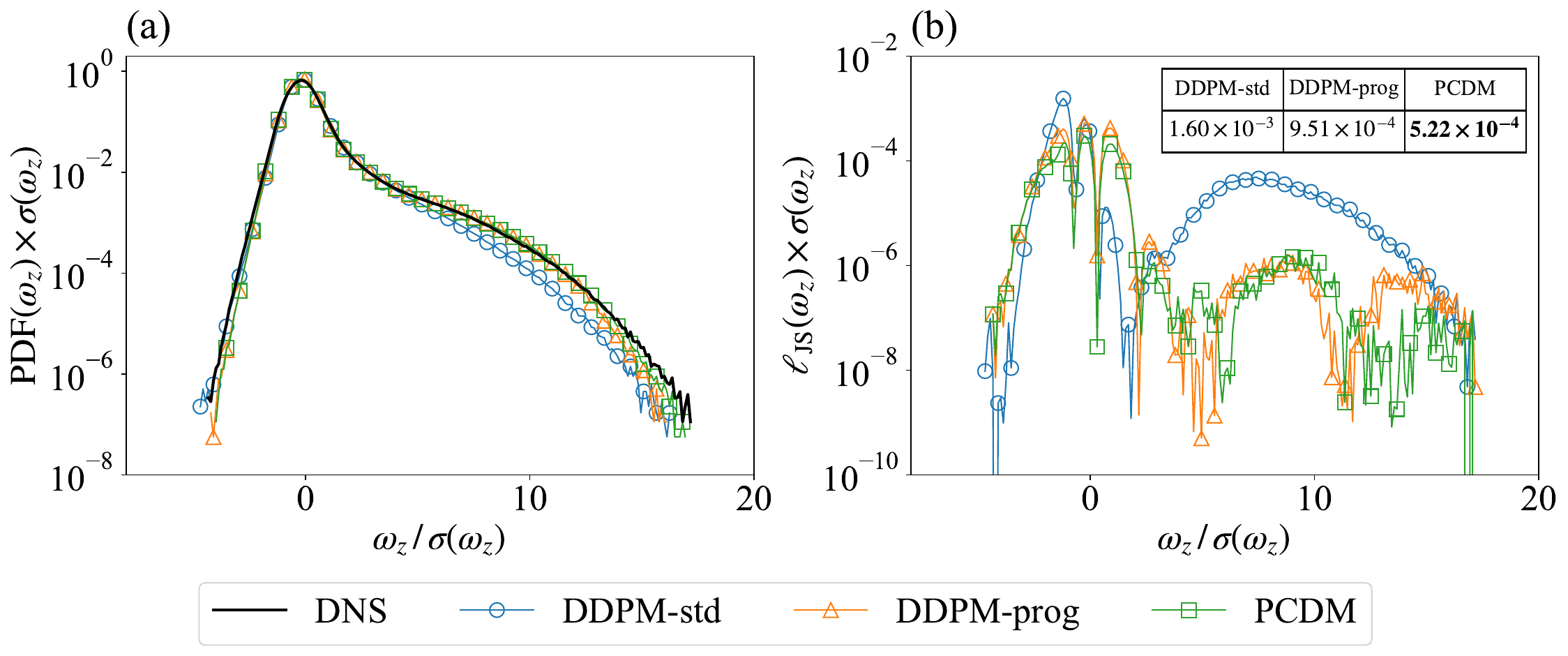}
\caption{\label{fig:vorticity_pdf}
\textbf{Single-point vorticity statistics.}
(a) PDFs of the vertical vorticity component $\omega_z$ for DNS, DDPM-std, DDPM-prog, and the PCDM, normalized by the DNS standard deviation $\sigma(\omega_z)$.
(b) Local contribution $\ell_{\mathrm{JS}}$ to the Jensen--Shannon divergence (JSD) between the model PDFs and the DNS PDF.
The inset table reports the integrated JSD $D_{\mathrm{JS}}$.}
\end{figure*}

Next, we assess the statistical fidelity of the generated fields using single-point vorticity statistics.
Figure~\ref{fig:vorticity_pdf}(a) compares the PDFs of the vertical vorticity component $\omega_z$ for DNS and the generated fields.
DDPM-std exhibits pronounced deviations in the distribution tails, systematically underestimating extreme vorticity events.
In contrast, both DDPM-prog and PCDM capture the heavy-tailed distributions characteristic of intermittent turbulence and show good agreement with the DNS.
To quantify the differences between the PDFs, Fig.~\ref{fig:vorticity_pdf}(b) reports the local contribution to the Jensen--Shannon divergence (JSD) between the model PDFs $p(x)$ and the DNS PDF $q(x)$, defined as
\begin{equation}
\ell_{\mathrm{JS}}(p(x)\|q(x))
=
\frac{1}{2}p(x)\log\frac{p(x)}{m(x)}
+
\frac{1}{2}q(x)\log\frac{q(x)}{m(x)},
\end{equation}
where $m(x) = \tfrac12 [p(x)+q(x)]$.
The JSD is then obtained as
\begin{equation}
D_{\mathrm{JS}}(p\|q) = \int \ell_{\mathrm{JS}}(p(x)\|q(x))\,dx.
\end{equation}
The divergence satisfies $0 \le D_{\mathrm{JS}} \le \ln 2$, where smaller values indicate better agreement between the distributions.
The inset table in Fig.~\ref{fig:vorticity_pdf}(b) reports the integrated JSD values for the different models.
PCDM yields the smallest divergence, indicating the closest agreement with the DNS statistics.
Consistent results are obtained for the horizontal vorticity PDFs (not shown).

\subsection{Divergence-free validation}

\begin{figure}[b]
\includegraphics[width=\linewidth]{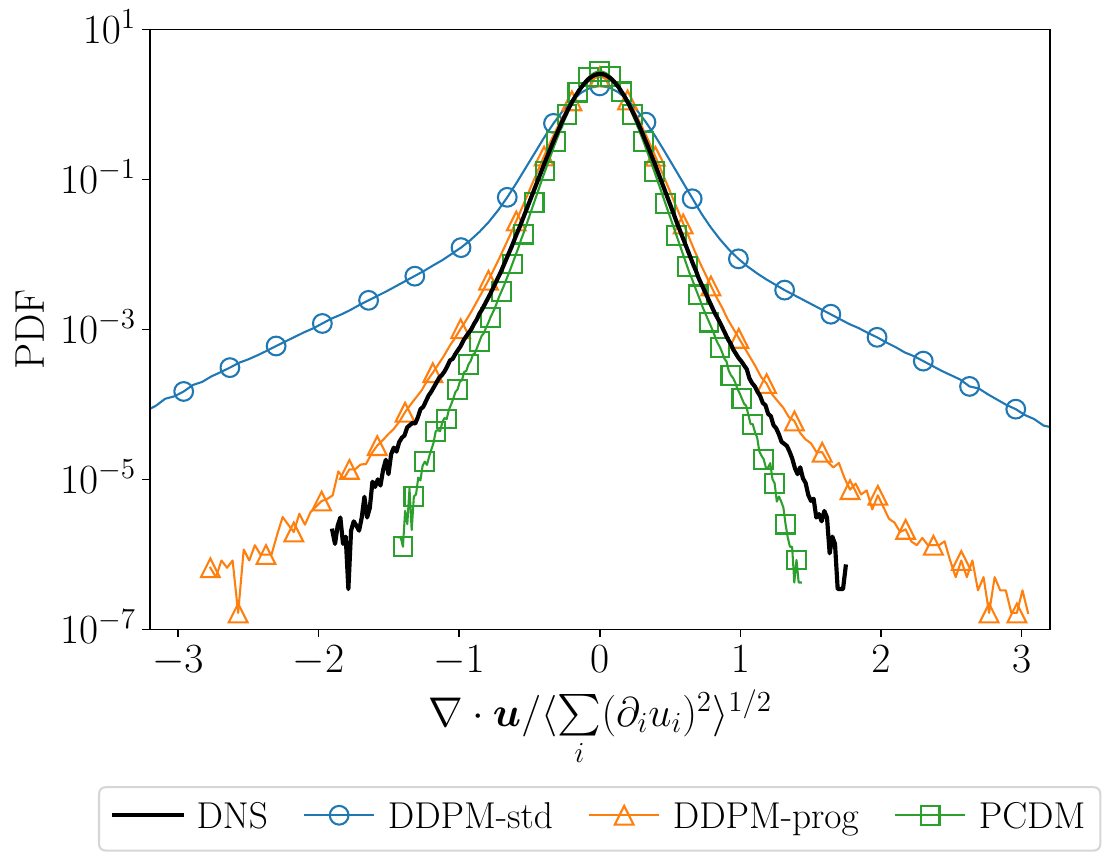}
\caption{\label{fig:incompressibility_pdf}
\textbf{Normalized velocity divergence probability density function.}
PDF of the normalized velocity divergence for DNS, DDPM-std, DDPM-prog, and the PCDM, computed using second-order finite differences.
}
\end{figure}

The velocity fields generated by the PCDM are inherently divergence-free at machine precision, as a direct consequence of the physics-constrained generative dynamics.
To quantify this property in a fair way, we compute a coarse-grained approximation of the gradients using second-order finite differences and compare the resulting divergence statistics among the three generative models considered here.
In Fig.~\ref{fig:incompressibility_pdf} we show the PDF of the normalized velocity divergence,
\[
\nabla \cdot \boldsymbol{u} \big/ \langle \sum_i (\partial_i u_i)^2 \rangle^{1/2},
\]
demonstrating that both DDPM-std and DDPM-prog exhibit larger deviations from zero, indicating larger compressibility errors than PCDM, which remains centered at zero and agrees with the DNS up to discretization accuracy.
The zero-mean momentum constraint is also violated by the unconstrained DDPMs, whereas both DNS and PCDM satisfy it to machine precision (not shown).

\subsection{\label{sec:robustness}
Robustness and convergence toward a consistent physical representation}

We next investigate the robustness of the PCDM framework.
Specifically, we examine whether distinct, non-equivalent implementations of
the same physical constraints give rise to similar generative behavior,
thereby assessing convergence toward a consistent underlying physical
representation.

To this end, we consider two variants of the PCDM. 
The first corresponds to the original formulation introduced in
Sec.~\ref{sec:rot_turb} and used throughout the results above,
and is denoted \textit{PCDM-Fourier}. 
In this variant, incompressibility and zero-mean flow are enforced exactly
through the Fourier-space projection operator, defined in
Eq.~(\ref{eq:projector_fourier}).

The second variant, referred to as \textit{PCDM-Integral}, employs a relaxed
formulation of the same physical constraints.
Instead of enforcing the exact pointwise incompressibility and zero-mean
constraints via the Fourier-space projection, we impose only a necessary
integral condition, requiring that the plane-integrated velocity components
vanish in each coordinate direction:
\[
\int u_i\,\mathrm{d}x_j\,\mathrm{d}x_k = 0,
\]
for the three cyclic permutations $(i,j,k) = (x,y,z)$, $(y,z,x)$, and
$(z,x,y)$.
This condition is weaker than the exact constraint in Eq.~(\ref{eq:constraints}); a field may satisfy it without being divergence-free at every point.
Accordingly, the Fourier-space projector $\mathcal{P}$ is replaced by an integral projector, $\mathcal{P}_{\mathrm{int}}$, which enforces
the above condition within the diffusion-based generative process.

Because the reverse process in Eq.~(\ref{eq:pcdm_reverse_process}) is stochastic,
we fix the random seed to generate paired velocity-field realizations from
PCDM-Fourier and PCDM-Integral using identical noise histories:
\[
\mathcal{V}^{(F)} \equiv \{\boldsymbol{u}^{(F)}(\boldsymbol{x}) \,|\, \boldsymbol{x} \in \mathcal{D}\},
\quad
\mathcal{V}^{(I)} \equiv \{\boldsymbol{u}^{(I)}(\boldsymbol{x}) \,|\, \boldsymbol{x} \in \mathcal{D}\}.
\]
We then quantify the similarity between paired realizations using the
cosine similarity:
\begin{equation}
\mathcal{C} =
\frac{\langle \boldsymbol{u}^{(F)}(\boldsymbol{x}) \cdot \boldsymbol{u}^{(I)}(\boldsymbol{x}) \rangle}
{\langle |\boldsymbol{u}^{(F)}(\boldsymbol{x})|^2 \rangle^{1/2}
 \langle |\boldsymbol{u}^{(I)}(\boldsymbol{x})|^2 \rangle^{1/2}},
\end{equation}
where $\langle \cdot \rangle$ denotes spatial averaging.

\begin{figure*}
\includegraphics[width=0.98\textwidth]{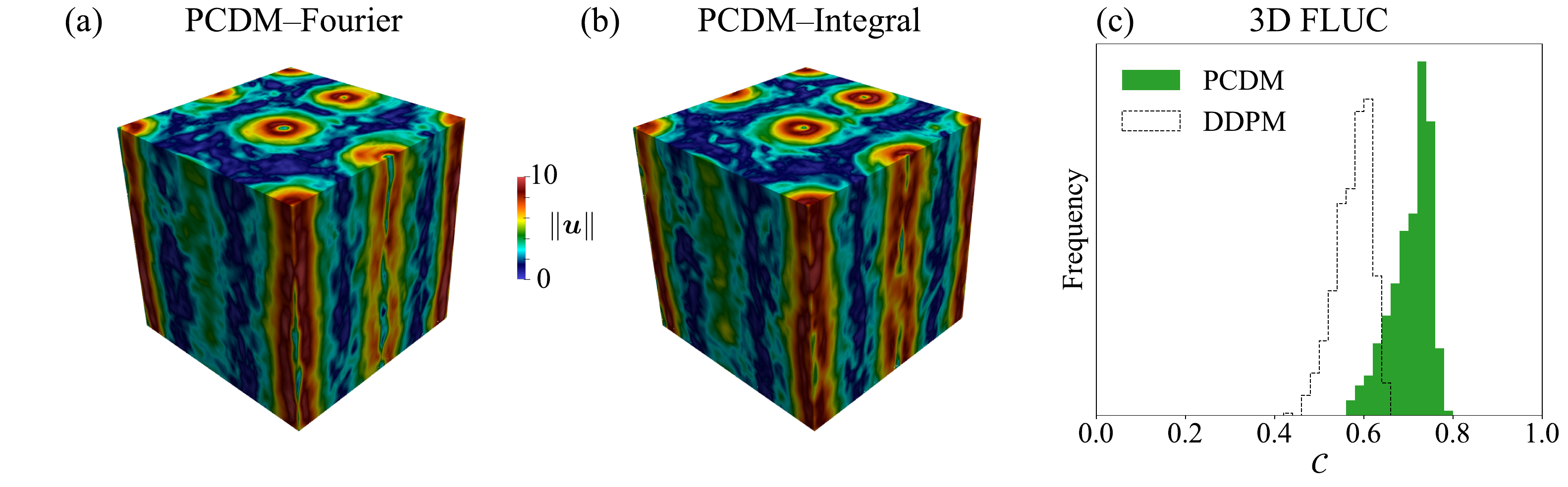}
\caption{\label{fig:robustness}
\textbf{Robustness of physics-constrained diffusion models under different constraint implementations.}
Paired samples are generated using identical noise histories during the backward diffusion process.
(a) Velocity magnitude of the full velocity field (dominated by the 2D3C component) generated using the Fourier-based constraint formulation (PCDM-Fourier).
(b) Corresponding field generated using the relaxed integral constraint formulation (PCDM-Integral). The two realizations are visually indistinguishable.
(c) Histogram of the cosine similarity $\mathcal{C}$ computed for the three-dimensional fluctuating component (3D FLUC) between paired samples.
In panel (c), the legend entry ``PCDM'' refers to the similarity between PCDM-Fourier and PCDM-Integral, while ``DDPM'' refers to the similarity between DDPM-prog and PCDM-Fourier (shown for reference).
The high similarity values indicate that the two PCDM variants generate highly correlated realizations even for the fully 3D fluctuations.
For the 2D3C component, both comparisons yield cosine similarity values concentrated near unity (not shown).
}
\end{figure*}

A representative pair of velocity fields generated by the two models is shown in Fig.~\ref{fig:robustness}(a,b).
The realizations are visually indistinguishable at the level of the full velocity field, which is dominated by the 2D3C component.
To quantify the similarity between the two PCDM variants, we compute the cosine similarity $\mathcal{C}$ for the 3D fluctuating component, which represents the most challenging part of the flow. While PCDM captures this component well, DDPM-prog fails to do so.
The resulting distribution of $\mathcal{C}$ is shown in Fig.~\ref{fig:robustness}(c).
Despite the physical non-equivalence of the two constraint formulations, the cosine similarity remains high, indicating that the two PCDM variants generate strongly correlated realizations when driven by the same stochastic noise.
This demonstrates that the PCDM framework is robust with respect to different implementations of the physical constraints and converges toward a consistent physical representation.

\section{\label{sec:conclusion}Conclusion}

In this work, we have shown that a direct application of standard denoising diffusion probabilistic models to fully developed three-dimensional turbulence does not achieve statistically faithful generation of velocity fields, even in the simplest single-snapshot setting.
The generated samples exhibit pronounced scale-dependent deviations in anisotropic energy spectra and intermittency statistics, together with systematic violations of fundamental physical constraints, including incompressibility and zero-mean momentum.
These failures persist despite apparent training convergence and the use of progressive training strategies, suggesting that they cannot be resolved by optimization improvements alone in systems with extremely large numbers of strongly coupled degrees of freedom.

This behavior points to a more fundamental limitation of standard diffusion training objectives when applied to complex physical systems governed by global constraints.
Because the denoising objective is blind to constraints such as incompressibility and zero-mean momentum, the learned distribution are not properly confined to physically admissible configurations, rendering modeling in such high-dimensional settings intrinsically fragile.

To address this limitation, we introduced a physics-constrained diffusion model (PCDM) designed to incorporate \emph{a priori} physical constraints directly into the learned generative distribution.
We demonstrated that the PCDM provides a stochastic generative modeling framework for 3D rotating turbulence at medium resolution that strictly satisfy the imposed physical constraints, while accurately reproducing anisotropic energy spectra and multiscale intermittency statistics, in close agreement with the DNS reference.

Our results open the way to new applications of generative models for data augmentation and data assimilation through guided sampling conditioned on observations, a problem of central importance in geophysical, environmental, and astrophysical contexts \cite{li2025stochastic, vishwasrao2025diff, giral2026genda, martin2025generation, amoros2026guiding}. In addition, such models may serve as physically consistent probabilistic priors for the statistical sampling of rare and extreme events in high-dimensional multiscale systems \cite{ragone2018computation}.

Although our work represents one of the first applications of generative diffusion models to fully 3D turbulence -- addressing important bottlenecks such as intermittency and multiscale structure, which are absent in 2D flows -- much work remains, both technologically and conceptually, before data-driven approaches can rival the capacity of the Navier-Stokes equations to generate fully developed turbulence at high resolution.
For example, no data-driven model to date is capable of generating three-dimensional turbulent flows at high Reynolds numbers for transformative applications. 
Such regimes require simulations at resolutions of billions (or even trillions) of grid points, with fully three-dimensional turbulent velocity fields evolved over long time horizons. We currently lack both high-quantity, high-quality ground-truth datasets for training and the necessary computational resources to achieve this goal.
Furthermore, it remains unclear whether the separation between generalization and memorization phases observed in diffusion models will persist for sufficiently large training datasets \cite{kadkhodaie2023generalization, bonnaire2025diffusion, achilli2025memorization}. Moreover, it is not yet known whether the remarkable accuracy demonstrated by diffusion models will remain satisfactory as the complexity of the target data increases, as occurs in fully developed turbulence, where the dimension of the turbulent attractor grows as $Re^{9/4}$, with $Re$ denoting the Reynolds number \cite{frisch1996turbulence}.

Beyond the specific setting considered here, our results point to a broader implication for diffusion-based generative modeling of complex physical systems.
In systems where admissible configurations are strongly constrained by physical laws, faithful generative modeling benefits from incorporating such constraints directly into the generative dynamics. Physics-constrained diffusion models thus offer a natural route toward aligning learned distributions with the structure of the physically admissible configuration space.
In this context, the present PCDM framework naturally invites extensions to other incompressible flow configurations, as well as to settings involving additional or dynamical constraints. However, the current formulation is tailored to constraints that act at the level of individual realizations and admit an explicit projection onto the admissible space. Extending this approach to constraints that are statistical, approximate, or only implicitly enforced remains an important direction for future work.

\section*{Data and Code Availability}

The data used to train the diffusion models in this study are available for
download from the open-access Smart-TURB portal at
\url{http://smart-turb.roma2.infn.it}, in the TURB-Rot repository
\cite{biferale2020turbrot}.
The code used to train the diffusion models and to generate samples from them
will be made publicly available upon publication.

\begin{acknowledgments}
This work was supported by the European Research Council (ERC) under the European Union’s Horizon 2020 research and innovation programme Smart-TURB (Grant Agreement No. 882340) and by the Italian Ministry of University and Research (MUR) - Fondo Italiano per la Scienza (FIS2) - 2023 Call, project DeepFL, CUP: E53C24003760001 and the FARE programme (No. R2045J8XAW).
The authors acknowledge CINECA for providing high-performance computing resources and support on the Leonardo supercomputer.
\end{acknowledgments}

\appendix

\section{\label{app:simulation}Numerical simulation and dataset preparation}

\subsection{Governing equations and simulation.}
We consider statistically stationary rotating turbulence governed by the three-dimensional incompressible Navier-Stokes equations in a periodic domain with uniform rotation,
\begin{equation}\label{eq:NS_rot}
    \frac{\partial \boldsymbol{u}}{\partial t}
    + \boldsymbol{u}\cdot\nabla \boldsymbol{u}
    + 2\boldsymbol{\Omega}\times\boldsymbol{u}
    = -\nabla \tilde{p}
    + \nu \Delta \boldsymbol{u}
    + \boldsymbol{f},
\end{equation}
where $\boldsymbol{u}$ is the 3D velocity field, $\tilde{p}$ is the modified pressure in the rotating frame enforcing incompressibility, $\nabla \cdot \boldsymbol{u} = 0$, and $\boldsymbol{\Omega} = \Omega \hat{\boldsymbol{z}}$ is the rotation vector. The flow is simulated using DNS with a pseudo-spectral method on a $256^3$ grid. 
A large-scale stochastic forcing $\boldsymbol{f}$ injects energy near the wavenumber $k_f = 4$, implemented as a second-order Ornstein-Uhlenbeck process~\cite{sawford1991reynolds, buzzicotti2016lagrangian}. 
The forcing scale is chosen to allow both forward and inverse energy cascades.
To extend the inertial range, the viscous term $\nu \Delta\boldsymbol{u}$ is replaced by a hyperviscous dissipation $\nu_h \nabla^{4}\boldsymbol{u}$, and a linear large-scale friction $\beta_f \Delta^{-1}\boldsymbol{u}$ is applied to suppress the formation of a large-scale condensate~\cite{alexakis2018cascades}. 
The flow reaches a statistically stationary regime characterized by a Rossby number $Ro = \mathcal{E}^{1/2}/(\Omega/k_f)\approx 0.1$, where $\mathcal{E} = \langle |\boldsymbol{u}|^2 \rangle/2$ is the kinetic energy. 
The integral length scale is defined as $L = L_0 / k_f \approx 0.25L_0$, and the corresponding eddy turnover time is $T_L = L/\sqrt{2\mathcal{E}} \approx 0.43$, where $L_0 = 2\pi$ is the domain size. Further details of the simulation are provided in Ref.~\cite{biferale2020turbrot}.

\subsection{\label{sec:rot_turb:dataset}Preparation of the training dataset.}

To enable generative modeling of 3D turbulent flow fields at single time instances, we construct a training dataset consisting of velocity snapshots sampled from the statistically stationary regime of the flow. 
A total of 600 snapshots are collected with a temporal spacing of $\Delta t_s \approx 2.34T_L$ between successive samples to ensure statistical decorrelation (see the inset of Fig.~\ref{fig:rotating_turbulence}(d)). 
The resolution of the sampled fields is then reduced from $256^3$ to $64^3$ by applying a Galerkin truncation in Fourier space,
\begin{equation}
    \boldsymbol{u}(\boldsymbol{x}) = \sum_{\lVert\boldsymbol{k}\rVert \leq k_\nu} \hat{\boldsymbol{u}}(\boldsymbol{k}) \, \mathrm{e}^{\mathrm{i}\boldsymbol{k}\cdot\boldsymbol{x}},
\end{equation}
where the cutoff wavenumber is set to the viscous Kolmogorov scale $k_\nu \approx 32$. 
This truncation removes only fully dissipative Fourier modes, corresponding to scales at which the flow dynamics are nearly linear, while retaining the entire inertial range and its nonlinear interactions. 
As a result, the reduced-resolution fields preserve all essential multiscale, anisotropic, and intermittent features of the flow, as well as the exact physical constraints of incompressibility and zero-mean momentum, while providing a computationally tractable representation for generative modeling.

\section{\label{app:spectral_definitions}Spectral definitions}

We define the kinetic energy spectrum as
\begin{equation}
    E(k) = \frac{1}{2} \sum_{k \le |\boldsymbol{k}'| < k+1} \sum_{i \in \{x,y,z\}} |\hat{u}_i(\boldsymbol{k}')|^2,
\label{eq:total_spectrum}
\end{equation}
where $\hat{u}_i(\boldsymbol{k}')$ denotes the Fourier coefficient of the $i$th velocity component $u_i(\boldsymbol{x})$ at the wavevector $\boldsymbol{k}'$.
The corresponding spectra of the 2D3C component and the 3D fluctuating field are denoted by $E_{2D3C}(k)$ and $E_{3D}(k)$, respectively, and are obtained from Eq.~\eqref{eq:total_spectrum} by using the Fourier coefficients of $\boldsymbol{u}_{2D3C}$ and $\boldsymbol{u}_{3D}$ defined in Sec.~\ref{sec:results}.

\section{\label{app:ddpm_details}
Forward process and reverse transition}

We briefly recall the forward diffusion process of DDPM.
The forward dynamics is defined as a Markov chain with Gaussian transitions
\begin{equation}
q(\mathcal{V}_t | \mathcal{V}_{t-1})
=
\mathcal{N}\big(
\mathcal{V}_t;
\sqrt{1-\beta_t}\mathcal{V}_{t-1},
\beta_t \mathbf I
\big),
\label{eq:forward_transition}
\end{equation}
where $\{\beta_t\in(0,1)\}_{t=1}^T$ is a prescribed noise schedule.
Marginalizing the Markov chain yields the closed-form expression
\begin{equation}
q(\mathcal{V}_t | \mathcal{V}_0)
=
\mathcal{N}\big(
\mathcal{V}_t;
\sqrt{\bar\alpha_t}\mathcal{V}_0,
(1-\bar\alpha_t)\mathbf I
\big),
\label{eq:forward_process_closed_form}
\end{equation}
where $\bar\alpha_t=\prod_{i=1}^{t}(1-\beta_i)$.

During training the clean sample $\mathcal V_0$ is available, and the conditional
posterior of the forward process can be written using Bayes' theorem as
\begin{equation}
q(\mathcal{V}_{t-1} | \mathcal{V}_t, \mathcal{V}_0)
=
\frac{
q(\mathcal{V}_t | \mathcal{V}_{t-1})
\, q(\mathcal{V}_{t-1} | \mathcal{V}_0)
}{
q(\mathcal{V}_t | \mathcal{V}_0)
}.
\end{equation}
Since all terms are Gaussian, the posterior admits the closed-form expression
\cite{ho2020denoising}
\begin{equation}
q(\mathcal{V}_{t-1} | \mathcal{V}_t, \mathcal{V}_0)
=
\mathcal{N}\big(
\mathcal{V}_{t-1};
\tilde{\mu}_t(\mathcal{V}_t, \mathcal{V}_0),
\tilde{\beta}_t \mathbf{I}
\big),
\end{equation}
with mean and variance given by
\begin{align}
\tilde{\mu}_t(\mathcal{V}_t,\mathcal{V}_0)
&=
\frac{\sqrt{\bar{\alpha}_{t-1}}\beta_t}{1-\bar{\alpha}_t}
\mathcal{V}_0
+
\frac{\sqrt{\alpha_t}(1-\bar{\alpha}_{t-1})}{1-\bar{\alpha}_t}
\mathcal{V}_t,
\\
\tilde{\beta}_t
&=
\frac{1-\bar{\alpha}_{t-1}}{1-\bar{\alpha}_t}
\beta_t.
\end{align}

Following \cite{ho2020denoising}, the reverse transition is parameterized
by replacing the unknown clean sample $\mathcal V_0$ with its model estimate
$\mathcal V_{0,\theta}$:
\begin{equation}
p_\theta(\mathcal{V}_{t-1}|\mathcal{V}_{t})
=
\mathcal{N}\big(
\mathcal{V}_{t-1};
\tilde{\mu}_t(\mathcal{V}_t, \mathcal V_{0,\theta}),
\tilde{\beta}_t \mathbf{I}
\big).
\label{eq:ddpm_step_approx}
\end{equation}

\section{\label{app:implementation}Implementation details}

\subsection{Neural network architecture}

\begin{figure*}
\includegraphics[width=\textwidth]{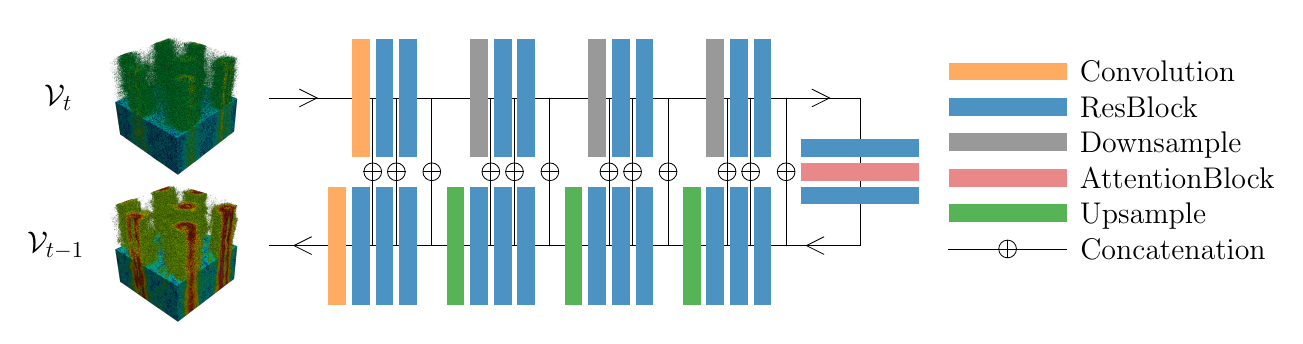}
\caption{\label{fig:architecture}
\textbf{Neural network architecture.}
Schematic of the U-Net architecture shared in all models, including DDPM-std, DDPM-prog, and the PCDM.
Different network modules are indicated in the legend.
}
\end{figure*}

All models considered in this work, including DDPM-std, DDPM-prog, and the PCDM, employ the same neural network architecture throughout all experiments.
The network adopts a U-Net architecture commonly used in diffusion models, adapted to 3D inputs, as illustrated in Fig.~\ref{fig:architecture}. The architecture consists of a contracting path and an expanding path, in which spatial resolution is reduced and restored in the two directions perpendicular to the rotation axis, while being preserved along the rotation axis. This defines four stages of spatial scales, at which sequences of residual blocks are applied with feature-channel widths $\{C, 2C, 4C, 8C\}$ and a base width $C=32$.
The contracting and expanding paths are connected at the bottleneck by an intermediate module consisting of two residual blocks and a four-head self-attention block. Time conditioning for the diffusion step is implemented by conditioning the network on the cumulative noise parameter $\bar{\alpha}_t$, following~\cite{chen2020wavegrad}.

\subsection{\label{app:progressive}Progressive training schedule}

We employ a progressive training schedule to alleviate the optimization difficulty
of standard DDPMs associated with the extremely high dimensionality of the full
3D velocity field.
In this procedure, the network architecture is kept fixed, which is possible
because the model operates convolutionally and does not depend on a fixed input
size, while the spatial support of the training data is gradually increased
along the rotation axis.

Specifically, training is initialized on velocity fields of size
$64\times64\times n_z$, where $n_z$ denotes the thickness along the rotation axis
(see Fig.~\ref{fig:training_convergence}(a)).
The value of $n_z$ is progressively increased as
$n_z = 8,\,16,\,32,\,48,$ and finally $64$, corresponding to the full domain.
At each stage, the model is trained until convergence before proceeding to the
next thickness.
Throughout this process, the network architecture and diffusion setup remain
unchanged, and only the spatial extent of the training data is increased.

\subsection{\label{app:training}Diffusion setup and training configuration}
All models employ the same diffusion setup with $T=2000$ diffusion steps and a
linear variance schedule, in which the noise variance increases linearly from
$\beta_1 = 10^{-6}$ to $\beta_T = 10^{-2}$.
In the reverse transition, we follow \cite{ho2020denoising} and fix the variance
to $\beta_t$ rather than $\tilde{\beta}_t$.
Optimization is performed using the AdamW optimizer with a learning rate of
$10^{-4}$.
All training and sampling are conducted on four NVIDIA A100 GPUs.

For DDPM-std and the PCDM, training is performed with a batch size of 64 for
$8\times10^5$ iterations, corresponding to approximately 12 days of training per
model.

For DDPM-prog, batch sizes of 256 and 128 are adopted for slab thicknesses
$n_z = 8,\,16$ and $n_z = 32$, respectively, while a batch size of 64 is used for
$n_z = 48$ and $64$.
The total number of training iterations across all stages is
$4\times10^6$, corresponding to a total training time of about 56 days.

During inference, an exponential moving average (EMA) of the model parameters
with a decay rate of 0.9999 is applied.
Generating 640 independent samples requires approximately 2 hours.


\bibliography{apssamp}

\end{document}